\documentclass[aps,prl,floatfix,twocolumn,tightenlines,amsmath,amssymb,nofootinbib]{revtex4-1}

\usepackage{graphicx}
\usepackage{amssymb}
\usepackage{color}
\usepackage{amsthm}
\usepackage{psfrag}
\usepackage{ifsym}
\usepackage{epstopdf}
\usepackage{dsfont}
\usepackage{amsmath,amsfonts}
\usepackage{multirow} 
\usepackage{mathtools}
\usepackage{graphicx}  
\usepackage{dcolumn}   
\usepackage{bm}        
\usepackage{enumerate}
\usepackage{array}
\usepackage{hyperref}
\usepackage{epsfig}

\newcommand{\bra}[1] {\langle #1 |}
\newcommand{\ket}[1] {| #1 \rangle}

\newcommand{\ketbra}[1]{ | #1 \rangle\!\langle #1 |}

\newcommand{\Tr} {\operatorname{Tr}}

\newcommand{\one}{\leavevmode\hbox{\small1\normalsize\kern-.33em1}}

\newcommand{\dket}[1]{\mbox{$\left|\left.#1\right\rangle\right\rangle$}}
\newcommand{\dbra}[1]{\mbox{$\left\langle\left\langle #1\right.\right|$}}
\newcommand{\dbradket}[2]{\mbox{$\langle\langle #1|#2\rangle\rangle$}}
\newcommand{\dketdbra}[2]{\mbox{$|#1\rangle\rangle\langle\langle #2|$}}

\newcommand{\be}{\begin{equation}}
\newcommand{\ee}{\end{equation}}
\newcommand{\beno}{\begin{equation*}}
\newcommand{\eeno}{\end{equation*}}
\newcommand{\bea}{\begin{eqnarray}}
\newcommand{\eea}{\end{eqnarray}}

\newcommand{\separator}{;}

\def\1#1{{\bf #1}}
\def\2#1{{\cal #1}}
\def\3#1{{\sl #1}}
\def\4#1{{\tt #1}}
\def\5#1{{\sf #1}}
\def\6#1{{\mathfrak #1}}
\def\7#1{{\mathbb #1}}

\newcommand{\id}{\mbox{$1 \hspace{-1.0mm} {\bf I}$}}

\definecolor{dred}{rgb}{.8,0.2,.2}
\definecolor{ddred}{rgb}{.8,0.5,.5}
\definecolor{dblue}{rgb}{.2,0.2,.8}
 
\begin{document}
\title{Characterizing quantum dynamics with initial system-environment correlations}

\author{M. Ringbauer$^{1,2\dagger}$, C.~J.~Wood$^{3,4}$, K.~Modi$^{5}$, A.~Gilchrist$^{6}$, A.~G.~White$^{1,2}$ and A.~Fedrizzi$^{1,2}$}
\affiliation{$^1$Centre for Engineered Quantum Systems, $^{2}$Centre for Quantum Computer and Communication Technology, School of Mathematics and Physics, University of Queensland, Brisbane, QLD 4072, Australia\\
$^{3}$ Institute for Quantum Computing, $^{4}$ Department of Physics and Astronomy, University of Waterloo, Ontario N2L 3G1, Canada\\
$^{5}$School of Physics, Monash University, VIC 3800, Australia\\
$^{6}$Centre for Engineered Quantum Systems, Department of Physics and Astronomy, Macquarie University, Sydney NSW 2113, Australia}

\begin{abstract}
\noindent We fully characterize the reduced dynamics of an open quantum system initially correlated with its environment.
Using a photonic qubit coupled to a simulated environment we tomographically reconstruct a superchannel---a generalised channel that treats preparation procedures as inputs---from measurement of the system alone, despite its coupling to the environment. We introduce novel quantitative measures for determining the strength of initial correlations, and to allow an experiment to be optimised in regards to its environment.
\end{abstract}

\maketitle

\begin{figure}[h!]
  \begin{center}
\includegraphics[width=\columnwidth]{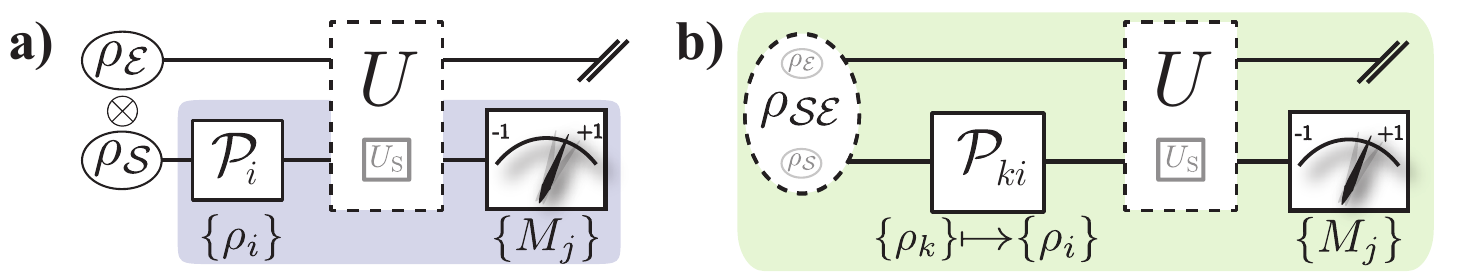}
  \end{center}
\vspace{-5mm}
\caption{System dynamics in the presence of an environment. \textbf{a)} Special case: with no initial $SE$ correlations the reduced evolution of the system, which interacts unitarily ($U$) with an environment, can be completely reconstructed from tomographically complete sets of input states $\{\rho_i\}$, prepared by preparation procedures $\{\2P_i\}$, and measurements $\{M_j\}$.
\\\textbf{b)} General case: the joint $SE$ state may be initially correlated before the state preparation procedure. The superchannel approach encompasses this situation by treating the preparation procedure $P_{i}$ as the input state to a more general description of the reduced system dynamics.}
  \label{fig:Motivation}
\end{figure}

\noindent In any real world experiment, quantum systems are inevitably coupled to their environment. This environment encapsulates all degrees of freedom that are not directly accessible to the experimenter and typically acts as a source of noise that needs to be constrained. Under certain conditions, however, it may also be harnessed as a resource---for example in initializing quantum states that may be otherwise unobtainable~\cite{Plenio1999,Bose1999,Beige2000,Diehl2008,Barreiro2011,Cormick2013,Lin2013,Xu2013,Wood2014}. 
In either case understanding the joint behaviour of system and environment is essential.
Quantum mechanics postulates that this joint system-environment ($SE$) state evolves unitarily, which need not be true for the system alone. Despite the coupling to the environment, the theory of open quantum systems allows for an operationally complete description of the reduced dynamics of the system, in the case that the initial $SE$ state is uncorrelated~\cite{Petruccione2002}, see Fig.~\ref{fig:Motivation}a).

In typical experiments, however, this central assumption is at best an approximation. In the presence of initial correlations, the procedure used to prepare a desired system input state may also affect the state of the environment. Any subsequent coupling between the system and environment can thus lead to drastically different reduced dynamics of the system conditional on the preparation~\cite{Modi2011}. Standard characterization techniques may in this case return a description of the reduced dynamics of the system that appears unphysical~\cite{Kuah2007,Ziman2006,Carteret2008,Modi2010,Modi2012a}. This highlights the importance of taking the environment into account to reliably characterize the system dynamics.

Recent results suggest that at least partial information about the initial joint $SE$ state can be extracted from measurements of the system alone. Initial correlations can be witnessed through the distinguishability~\cite{Wissmann2013,Laine2010,Rodriguez-Rosario2012,Gessner2011} and purity~\cite{Rossatto2011} of quantum states, which has also been explored experimentally~\cite{Smirne2011,Li2011,Gessner2013}. A more operationally complete characterisation can be obtained by explicitly treating the system's preparation procedure, rather than the resulting prepared state, as the input to the reduced description~\cite{Modi2012a}. 

This \emph{superchannel} approach captures not just the system evolution, but also the dynamical influence of the environment, even in the presence of initial $SE$ correlations. Here we present the first experimental demonstration of this technique by characterizing the evolution of a photonic qubit that is initially correlated with a simulated single-photon environment. We introduce novel quantitative measures for the strength of initial system-environment correlations and for optimizing experiments coupled to an environment. Finally, we develop maximum likelihood methods that can be applied to this formalism to overcome the issue of experimental noise.

The theory of open quantum systems provides a mathematically rigorous framework for the treatment of the dynamics of a quantum system $S$ interacting with an environment $E$, under the assumption that these are initially uncorrelated. 
The state of a $d$-dimensional open quantum system is described by a density matrix $\rho$, a positive semidefinite operator with trace one from the set of square matrices $L(\mathcal X)$, acting on the Hilbert space $\mathcal X\cong \mathbb{C}^d$. 
The evolution of open quantum systems is most generally described by a \emph{channel} $\mathcal{E}\colon L(\mathcal{X}_1)\to L(\mathcal{X}_2)$ which is a completely positive (CP) linear map from operators on $L(\mathcal X_1)$ to operators on $L(\mathcal{X}_2)$. In the following we assume that $\mathcal X_1{=}\mathcal{X}_2$, and keep the subscripts to distinguish between input and output Hilbert spaces, though the following results apply equally when the input and output spaces are of different dimensions.

A map $\mathcal{E}$ is positive if it preserves an operator's positivity and CP if the same is true for the composite map $\mathcal{I}\otimes\mathcal{E}$, where $\mathcal{I}$ is the identity map on a space at least as large as $\2X_1$. Any CP-map $\mathcal{E}$ is completely characterized by its Choi matrix $\Lambda_{\mathcal{E}}$, a positive-semidefinite operator $\Lambda_{\mathcal{E}}\in L(\mathcal{X}_1\otimes\mathcal{X}_2)$~\cite{Choi1975}. The Choi matrix may be constructed via the Choi-Jamio{\l}kowski isomorphism~\cite{Jamiokowski1972} via $\Lambda_{\mathcal{E}} = \sum_{ij}\ket{i}\bra{j}\otimes \mathcal{E}(\ket{i}\bra{j})$, where $\{\ket{i}\}_{i=0}^{d-1}$ is an orthonormal basis for $\2X_1$, and the evolution of the system state $\rho$ is given by $\mathcal{E}(\rho) = \Tr_{\2X_1}[(\rho^T\otimes\id)\Lambda_{\mathcal{E}}]$. 

The Choi matrix of an unknown quantum process can be reconstructed through quantum process tomography (QPT) from the outcomes of a finite set of measurements $\{M_j\}_{j=1}^{d^2}$, performed on a finite set of input states $\{\rho_i\}_{i=1}^{d^2}$ for the quantum system, see Fig.~\ref{fig:Motivation}a). Crucially, this assumes that the channel $\mathcal{E}$ being characterized is independent of the system's preparation.

In the presence of initial correlations between the system and the environment, this assumption is in general not satisfied. The joint system-environment state is then $\rho_{\textsc{se}}\in L(\mathcal{X}\otimes\mathcal{Y})$, as illustrated in Fig.~\ref{fig:Motivation}b), where $\mathcal{X}$ and $\mathcal{Y}$ are the state spaces of the system and environment, respectively. In the first step of the experiment the system is prepared in the state $\rho_S$ by applying a preparation map $\mathcal P$ to $S$ alone. Such a preparation will typically leave the environment in a state conditional on $\mathcal P$, which in turn leads to conditional dynamics $\mathcal{E}_{\mathcal{P}}$ and QPT would return a map which is a combination of the partial reconstructions of the possible $\mathcal{E}_{\mathcal P}$. In the following we consider the case of a fully de-correlating preparation procedure: $(\mathcal{P}\otimes\mathcal{I})(\rho_{SE}) = \rho_S\otimes \rho_{E\mid \mathcal{P}}$ where $\mathcal{I}$ is the identity map on $E$.
Denoting by $U$ the channel that describes the subsequent evolution of the system and environment, the final output state is given by $\rho_S^\prime = \Tr_E\left[ U\big((\mathcal P\otimes\id)(\rho_{\textsc{se}})\big)\right]$.

To characterize the system in the presence of possible initial correlations we describe the dynamics by means of a superchannel $\mathcal{M}\colon \mathcal{P} \rightarrow \rho_S^\prime$. $\mathcal{M}$ is a CP-map which takes as its input the chosen preparation procedure. While $\2M$ may be thought of as a regular channel with the input Hilbert space $L(\2X\otimes\2X)$ rather than $L(\2X)$, we prefer the terminology superchannel to emphasize the fact that it takes the preparation channel as an input, rather than the prepared state. The output is then given by $\rho_S^\prime = \mathcal{M}(\Lambda_{\mathcal{P}})$, where $\Lambda_{\mathcal{P}}$ is the Choi matrix for the preparation map $\mathcal{P}$~\cite{Modi2012a}.

\begin{figure}[h!]
  \begin{center}
  \includegraphics[width=\columnwidth]{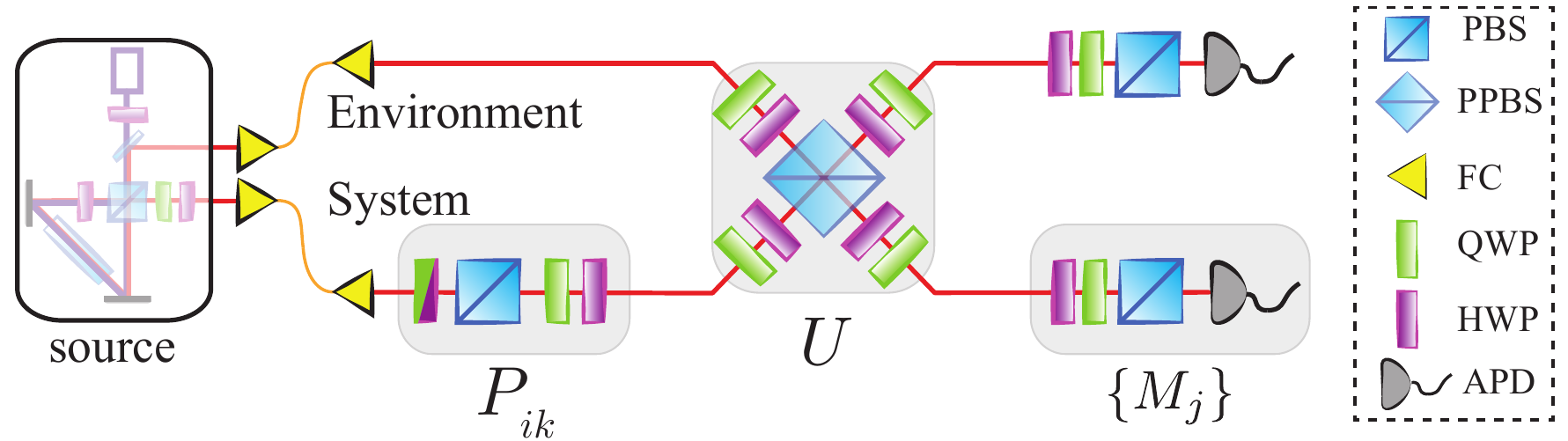}
  \end{center}
  \vspace{-5mm}
\caption{Experimental setup. System and environment photons are created with a controllable degree of entanglement between them, using the source of Ref.~\cite{fedrizzi2007wtf}. 
A combination of quarter- and half-wave plates (QWP, HWPs), and a polarizer (PBS) implement arbitrary preparations $\mathcal{P}_{ik}$ on the system. Both qubits are then subjected to a CZ gate between a set of HWPs and QWPs which simulates the environmentally-coupled single-qubit unitary $U$. This gate is based on non-classical interference at a partially polarizing beam splitter (PPBS) with reflectivities of $r_H=0$ ($r_V=2/3$) for horizontally (vertically) polarized light~\cite{Langford2005}. Finally, projective measurements on both qubits are performed by means of a combination of QWP, HWP and PBS before being detected by a single photon detector (APD).}
  \label{fig:setup}
\end{figure}

We now demonstrate this method by characterising the evolution of a single photonic qubit, coupled to a simulated environment as shown in Fig.~\ref{fig:setup}. Modelling this environment as a system of the same size is sufficient to describe a large range of dynamics and to illustrate the technique~\cite{Smirne2011,Chiuri2012}. However, a slightly larger environment would be required in the most general case~\cite{Schumacher1996,Narang2007}.
The initial $SE$ state was generated via spontaneous parametric down conversion and engineered to have the form
\begin{equation}
\ket{\psi}_{SE} = \cos (2\theta) \ket{H}_S\ket{V}_E + \sin (2\theta)\ket{V}_S\ket{H}_E ,
\end{equation} 
where $\ket{H},\ket{V}$ correspond to horizontally and vertically polarized photons respectively. In this case the strength of the initial correlations (both quantum and classical) is parametrized by the tangle $\tau = \sin^2(4\theta)$ and can be tuned from uncorrelated ($\tau=0$) to maximal correlation ($\tau = \pi/8$)~\cite{fedrizzi2007wtf}. 
We prepared states with varying degree of correlation, $\tau = \{ 0.012, 0.136, 0.423, 0.757, 0.908\}$, with an average fidelity of $F=0.96(1)$ with the corresponding ideal state. The system was then subjected to the preparation procedure $P_{ij}$, which prepared it in the state $\rho_j$ by first projecting onto the state $\rho_i$ followed by a unitary rotation. Here the indices $i,j\in \{\ket H,\ket V,\ket D,\ket A,\ket R,\ket L\}$, where $\ket{D/A}=\left(\ket H \pm \ket V\right)/\sqrt{2}$ and $\ket{R/L}=\left(\ket H \pm i \ket V\right)/\sqrt{2}$.

Imagine the experimenter aims to implement the target system unitary $U_{\textsc{s}}$ as $Z$, Hadamard $H=RZR^\dagger$, or $R_{\textsc{y}}Z$ which is a $\pi/4$-rotation around $\sigma_y$. The environmental coupling is simulated by replacing $Z$ by a controlled-Z (CZ) gate, which allows the environment to change the applied system unitary. Conditional on the state of the environment, the $Z$ part in the system evolution might be switched off. In the case of $Z$ and $H$, this results in the failure of the system unitary (i.e.\ the identity operation is implemented), while in the case of $R_{\textsc{y}}$ the environment can introduce a phase error. 

\begin{figure}[t!]
  \begin{center}
\includegraphics[width=\columnwidth]{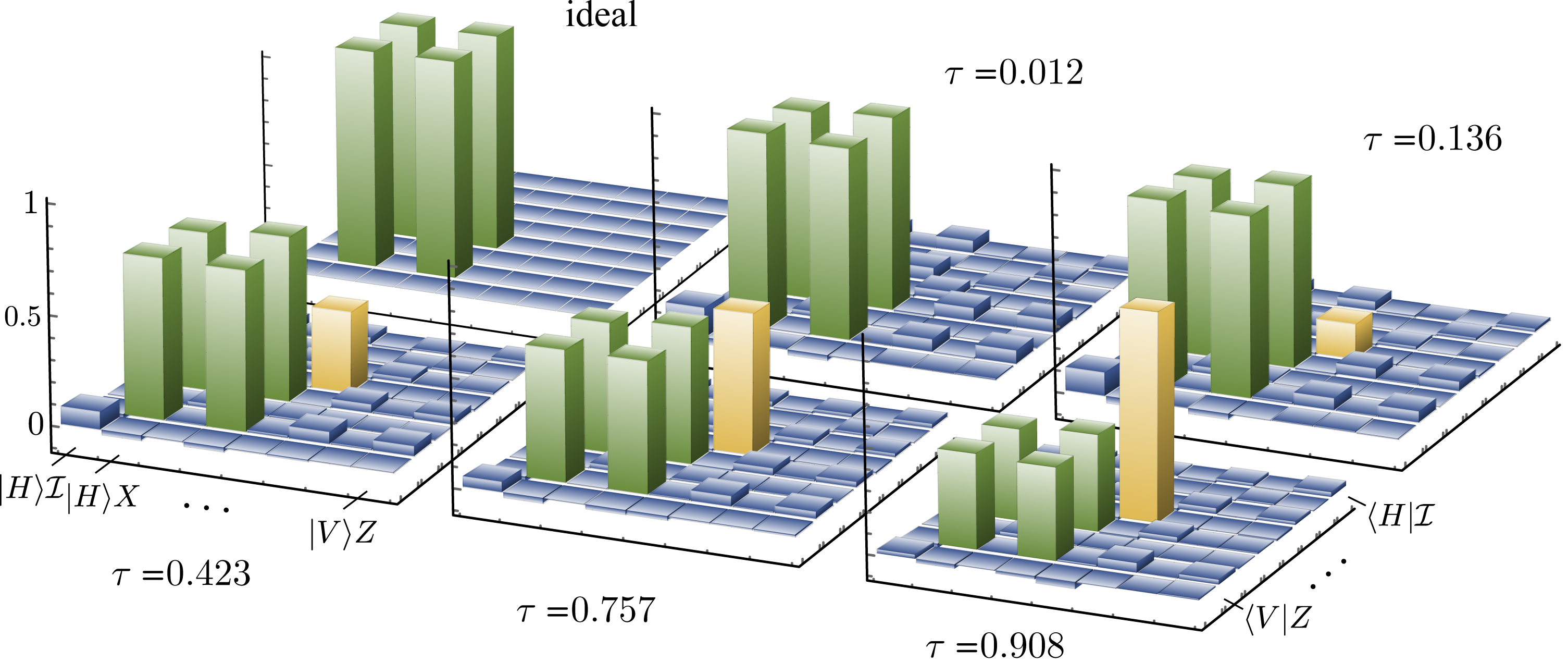}
  \end{center}
  \vspace{-5mm}
\caption{Real parts of $\Lambda_{\mathcal{M}}$ for a nominal $U_{\textsc{s}}=H$ operation on the system in the ideal, uncorrelated case and measurement results for increasing strength of initial correlations. The matrices $\Lambda_{\mathcal M}$ are shown in a polarization-Pauli basis, with the elements from left to right corresponding to $\{\ket{H},\ket{V}\}\otimes\{\mathcal{I},\sigma_{\textsc{x}},\sigma_{\textsc{y}},\sigma_{\textsc{z}}\}$ and from front to back corresponding to $\{\bra{H},\bra{V}\}\otimes\{\mathcal{I},\sigma_{\textsc{x}},\sigma_{\textsc{y}},\sigma_{\textsc{z}}\}$.
A common feature of the simulated interaction is the emergence of a peak corresponding to the identity operation (shown in yellow), confirming that the single-qubit operation $U_{\textsc{s}}$ (shown in green) has an increasing tendency to fail in the presence of stronger initial correlations. The negligible imaginary parts are not shown.}
  \label{fig:results_map}
\end{figure}

From measurements of the system along $\{H,V,D,A,R,L\}$, we can reconstruct the operator $\mathcal{M}$ via linear inversion as described in \cite{Modi2012a}. To avoid known problems with this technique, we used maximum likelihood estimation to enforce the reconstruction to be CP, see \ref{sec:SI_MLE}. For $U_{\textsc{s}}=H$ the reconstructed Choi matrix $\Lambda_{\mathcal{M}}$ is illustrated in Fig.~\ref{fig:results_map}. The maps for all other implemented $U_{\textsc{s}}$ are shown in Fig.~\ref{fig:results_CZ}-\ref{fig:results_CR_Im}. 
To allow for an operational interpretation of $\Lambda_{\mathcal{M}}$, it is displayed using the polarization basis for the index corresponding to the effective initial state, and the Pauli basis for the indices corresponding to the effective channel. With increasing strength of initial correlations we observe the emergence of an additional peak in the real part of $\Lambda_{\mathcal{M}}$, which is characteristic for the simulated $SE$ coupling.

The quantum superchannel $\mathcal M$ contains information about initial SE correlations that are visible through their effect on the subsequent experiment.
For any $\mathcal M$ we define an \emph{average initial system state} $\rho_{S,\text{av}} {=} \Tr_{23}[\Lambda_{\mathcal{M}}]/d$ and an \emph{average effective map} for the evolution of the system as $\Lambda_{\mathcal{E}_{\text{av}}} {=} \Tr_1[\Lambda_{\mathcal{M}}]$. 
For a simply separable initial state ($\rho_{SE}{=}\rho_S\otimes \rho_E$) the map $\mathcal M$ takes the simply separable form $\Lambda_{\mathcal{M}} {=} (\rho_S\otimes \Lambda_{\mathcal{E}})$. In this case $\rho_{S,\text{av}}{=}\rho_S$, and $\Lambda_{\mathcal{E}_{\text{av}}} {=} \Lambda_{\mathcal{E}}$ is the Choi matrix of the channel $\2E$ describing the (noisy) evolution of the system alone---the same as would result from conventional QPT. 
For a given $\mathcal M$ we can now define the corresponding separable superchannel $\mathcal{M}_s$ via $\Lambda_{\mathcal{M}_s} = (\rho_{S,\text{av}}\otimes\Lambda_{\mathcal{E}_{\text{av}}})$. In general $\mathcal M \neq \mathcal{M}_s$ and the distance between $\mathcal{M}$ and $\mathcal{M}_s$ can be used to quantify the strength of the initial $SE$ correlations. We thus define the \emph{initial correlation norm}:
\begin{equation}
\|\mathcal{M}\|_{\textsc{ic}} = \frac 1 2 \| \mathcal{M} - \mathcal{M}_s \|_{\diamondsuit} .
\label{eq:InitialCorrNorm}
\end{equation}

The matrix $\mathcal{M}-\mathcal{M}_s$ was introduced as \emph{correlation memory matrix} in Ref.~\cite{Modi2012a} since it describes how the dynamics is affected by initial correlations. Our choice of the diamond norm $\| \cdot \|_{\diamondsuit}$~\cite{Kitaev1997} allows for an operational interpretation of the IC-norm in terms of channel discrimination~\cite{Rosgen2005}. For any two quantum channels $\mathcal{E}_1,\mathcal{E}_2$, the best single shot strategy for deciding if a given channel is $\mathcal{E}_1$ or $\mathcal{E}_2$ succeeds with probability $\frac 1 2 \left( 1 + \frac 1 2 \| \mathcal{E}_1 - \mathcal{E}_2 \|_{\diamondsuit} \right)$. Thus when $\|\mathcal{M}\|_{\textsc{ic}}=0$, we are unable to operationally distinguish $\mathcal{M}$ and $\mathcal{M}_s$ which means that the system has no observable correlations with the environment. This can either mean that the initial $SE$ state is indeed uncorrelated, or that the environment is Markovian and initial correlations do not affect the subsequent dynamics. The initial correlation norm thus provides a necessary and sufficient condition for the decoupling of the future state of the system from its past interactions with the environment.
When $\|\mathcal{M}\|_{\textsc{ic}} > 0$ there exists an optimal preparation procedure that can be used as a witness for initial correlations, and the specific value of the norm determines the single shot probability of success for this witness. 

\begin{figure}[h!]
  \begin{center}
\includegraphics[width=\columnwidth]{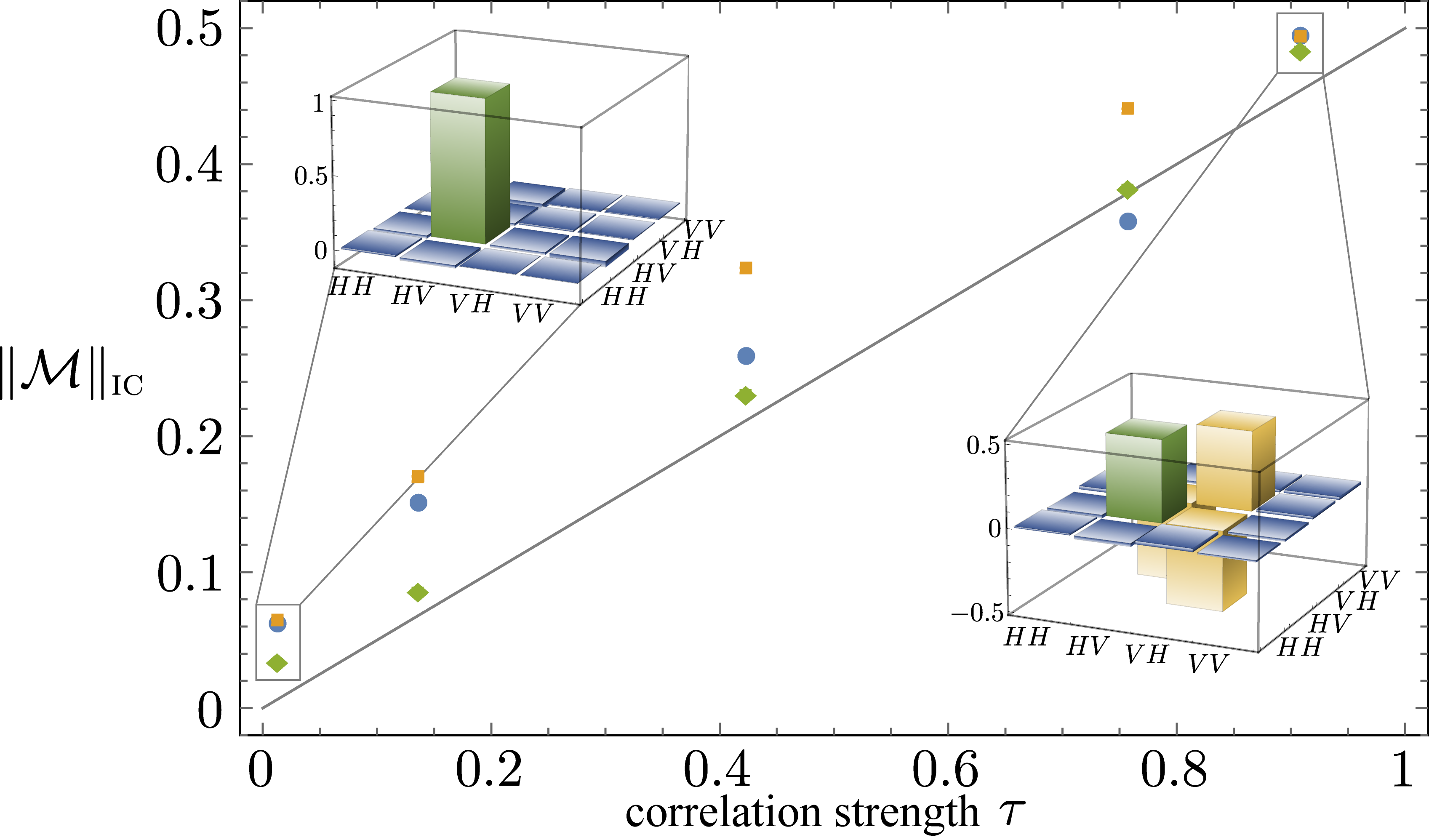}
  \end{center}
  \vspace{-5mm}
\caption{Initial correlation norm $\|\mathcal{M}\|_{\textsc{IC}}$ vs correlation strength $\tau$ of the initial $SE$ state for $U=\sigma_{\textsc{z}}$ (blue circles), $U=H$ (yellow squares) and $U=R_{\textsc{y}}$ (green diamonds). The values of $\tau$ were obtained from quantum state tomography of the initial system-environment state used for each experiment. The real parts of the states with weakest and strongest initial correlations are shown in the respective insets; imaginary parts are smaller by at least one order of magnitude and thus not shown. The solid line corresponds to the IC-norm in the ideal case. Error bars from Poissonian counting statistics are on the order of the symbol size.}
  \label{fig:results_correlation}
\end{figure}

Our measurements of $\|\mathcal{M}\|_{\textsc{ic}}$, plotted against the correlation strength $\tau$ of the simulated initial $SE$ states are shown in Fig.~\ref{fig:results_correlation}. For all three $SE$ interactions $U$ the maximum obtained value of $\|\mathcal{M}\|_{\textsc{ic}}$ is approximately $0.5$, which is in agreement with theoretical expectations.

The information contained in $\mathcal M$ can further be used to optimize the impact of the environment. To illustrate this, we introduce the measure of \emph{preparation fidelity} ${F_{\text{prep}}}$ for the case where high-fidelity projective preparation procedures are readily available, such as in photonic experiments.
Consider the scenario where the system is prepared via initial post-selection on the state $\rho_1$. The subsequent evolution is then described by the effective map $\mathcal{E}_{\rho_1}$ given by
\begin{equation}
\Lambda_{\mathcal{E}_{\rho_1}} = \frac{1}{p_{\rho_1}} \Tr_{1}\left[(\rho_1^\dagger\otimes\id_{23}) \Lambda_{\mathcal{M}}\right],
\label{eq:effChannel}
\end{equation}
where $p_{\rho_1} = \Tr\left[(\rho_1^\dagger\otimes\id_{23}) \Lambda_{\mathcal{M}}\right]/d$ is the probability of success for the post-selection on $\rho_1$. 
Studying these effective maps for different $\rho_1$ allows us to find the optimal preparation procedure for any desired evolution of the system. 
The measure ${F_{\text{prep}}}$ quantifies the process fidelity between the implemented effective map $\mathcal{E}_{\rho_1}$ and the desired target channel $\mathcal U$ for an initial projection onto the state $\rho_1$,
\begin{equation}
F_{\text{prep}}(\mathcal{M},\rho_1,U_{\textsc{s}}) = \frac{1}{d^2}F(\Lambda_{\mathcal{E}_{\rho_1}},\Lambda_{U_{\textsc{s}}}).
\label{eq:AvgPrepFidelity}
\end{equation}
The average preparation fidelity over all initial projections can be obtained from $\overline{\2E_{\rho_1}} = \2E_{av}$ defined previously by $\Tr_1[\Lambda_{\2M}]$.
On the other hand, maximizing $F_{\text{prep}}$ over all states ${\rho_1}$ for a given target unitary $U_{\textsc{s}}$ finds a preparation which allows for the highest quality implementation of the target unitary. Note that this is not equivalent to minimizing the impact of the environment, since the optimal preparation might harness some of the environmental correlations to improve the gate performance.

\begin{figure}[h!]
  \begin{center}
\includegraphics[width=\columnwidth]{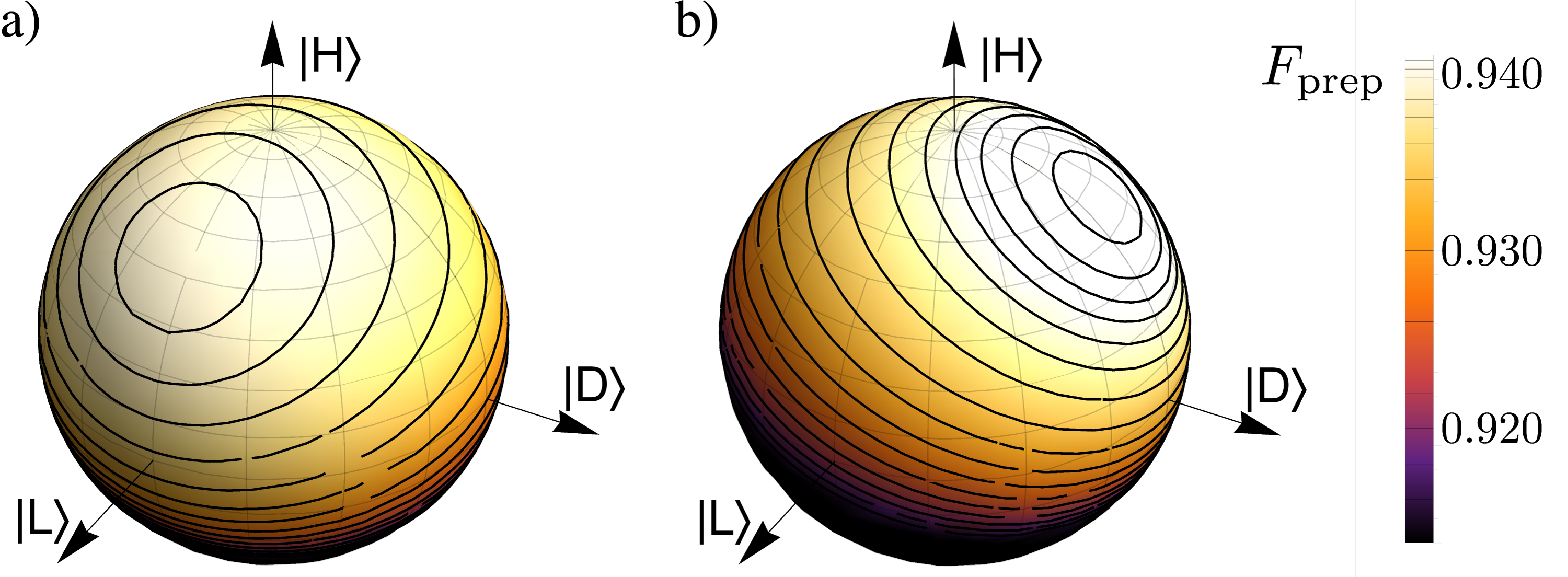}
  \end{center}
  \vspace{-5mm}
  \caption{Optimization of the preparation procedure. The average preparation fidelity $F_{\text{prep}}(\mathcal{M},\rho_1,Z)$ for \textbf{a)} $U_{\textsc{s}}=Z$ and \textbf{b)} $U_{\textsc{s}}=R_{\textsc{y}}Z$ is shown as a density plot on the surface of the Bloch sphere of the initial-projection state $\rho_1$. In both cases, we chose the lowest strength of initial correlation realized in the experiment to visualize the effect even for very weak SE correlations.}
  \label{fig:results_optimization}
\end{figure}

We now use our experimentally obtained $\mathcal{M}$ to optimise for maximum fidelity for the target $U_{\textsc{s}}=Z$, Fig.~\ref{fig:results_optimization}a), and $U_{\textsc{s}}=R_{\textsc{y}}Z$, Fig.~\ref{fig:results_optimization}b), given a correlation strength of $\|\mathcal M\|_{\textsc{ic}}=0.075(5)$ and $\|\mathcal M\|_{\textsc{ic}}=0.067(4)$, respectively. In the case shown in Fig.~\ref{fig:results_optimization}a), the effect of the environment is minimized for initial projection onto the state $\cos[\theta] \ket{H} + e^{i \varphi}\sin[\theta] \ket{V}$ with $\theta\approx 0.658$ and $\varphi\approx 0.252$. This demonstrates, that even for nearly uncorrelated $SE$ states, the chosen preparation procedure affects the achieved fidelity. In this example, the projection on the optimal state instead of the basis state $\ket{H}$ improved the fidelity by $0.2\%$. Similarly, minimizing $F_{\text{prep}}$ finds the worst-case preparation, which could give insight into where and why an experimental setup fails.

The conditions leading to CP reduced-dynamics of the system have received considerable interest~\cite{Rodriguez-Rosario2008,Shabani2009,Brodutch2013,Dominy2013}, however the known theoretical results are not particularly experimentally amenable. In contrast, the superchannel approach is operationally significant and experimentally accessible as demonstrated here. 
Notably, the reconstruction of $\mathcal M$ is a direct generalization of QPT, based on subjecting the system to $d^4$, rather than $d^2$ linearly-independent preparation procedures $\mathcal{P}_{ij}$, which need not all be fully de-correlating. Therefore, all the tools developed to improve the efficiency of QPT, such as compressive sensing~\cite{Shabani2011, Flammia2012} can also be applied to the reconstruction of $\mathcal{M}$.
 
Since the output of a channel is determined solely by the input and the channel itself, it can be thought of as a Markovian two-point connection. The presence of initial correlations, however, demands the use of the superchannel approach, which is thus a first step towards operationally and experimentally understanding non-Markovian quantum processes. Along these lines, the superchannel approach has also recently been used to derive the lower bound on entropy production in a generic quantum process~\cite{Vinjanampathy2014}. 
 
Our technique is most useful to quantum architectures which are strongly coupled to their environment, such as spins in local spin baths. Another application is in quantum control, where control timescales can be much faster than environmental reset times. Finally, it has been suggested that non-Markovianity can be exploited as a resource~\cite{bylicka2013non}; we showed how the superchannel formalism can be used to that extent in our gate optimisation.

\begin{acknowledgments}
KM thanks M.-H. Yung and C. A. Rodr\'iguez-Rosario and AF thanks B.P. Lanyon for helpful discussions. This work was supported in part by the Centres for Engineered Quantum Systems (CE110001013) and for Quantum Computation and Communication Technology (CE110001027). CJW acknowledges support through the Canadian Excellence Research Chairs (CERC) program and the Collaborative Research and Training Experience (CREATE) program, AGW through a UQ Vice-Chancellor's Senior Research Fellowship, and AF through an ARC Discovery Early Career Researcher Award, DE130100240
\end{acknowledgments}

\onecolumngrid
\clearpage
\renewcommand{\theequation}{S\arabic{equation}}
\renewcommand{\thefigure}{S\arabic{figure}}
\renewcommand{\thetable}{\Roman{table}}
\renewcommand{\thesection}{S\Roman{section}}
\setcounter{equation}{0}
\setcounter{figure}{0}
\begin{center}
{\bf \large Supplemental Material}
\end{center}
\twocolumngrid

Here we discuss in more detail the reconstruction of the superchannel $\mathcal M$ from experimental data following a direct, linear-inversion approach, as well as using more sophisticated maximum likelihood estimation to account for errors from statistical fluctuations. We also present additional experimental data. 
In the final section we present graphical techniques in terms of \emph{tensor networks}, which greatly simplify the derivation and manipulations of $\2 M$

\section{Conventions}
\label{Sec:Supp1}
\subsection{Choi matrix representation}
Let $\2X\cong \mathbb C^d$ be a $d$-dimensional complex Hilbert space, $L(\2X)$ represent the Hilbert space of $d\times d$ complex matrices $A: \2X\rightarrow \2X$, and $T(\2X_1,\2X_2)$ represent the Hilbert space of operator maps $\2E: L(\2X_1)\rightarrow L(\2X_2)$. 

An operator map $\2E\in T(\2X_1,\2X_2)$ may be completely described by the Choi matrix $\Lambda_{\2E}\in L(\2X_1\otimes\2X_2)$ which is constructed via the Choi-Jamio{\l}kowski isomorphism
\beno
\Lambda_{\2E } = \sum_{i,j=0}^{d-1}  \ket{i}\bra{j}\otimes \2E(\ket{i}\bra{j}).
\eeno
$\Lambda_{\2E}$ is a positive semidefinite matrix if and only if $\2E$ is a completely positive (CP) map, $\Tr_2[\Lambda_{\2E}]=\id$ if and only if $\2E$ is a trace preserving (TP) map. Further it is normalized by $\Tr[\Lambda_{\2E}] = \Tr[\2E(\id)]$, that is $\Tr[\Lambda_{\2E}]=d$ for a TP map.  

The evolution of state $\2E$ is then described by
\be
\2E(\rho) = \Tr_{\2X_1}[(\rho^T\otimes\id)\Lambda_{\2E}]
\label{eq:choievo}
\ee
where $\Tr_{\2X_1}$ is the partial trace over $L(\2X_1)$, and $^T$ denotes transposition.

In terms of index contraction we may rewrite Eq.~\eqref{eq:choievo} as
\beno
\2E(\rho)_{i \separator j} = \sum_{n,m} (\rho)_{n \separator m} (\Lambda_{\2E})_{ni \separator mi}
\eeno
where the tensor components are given by
\begin{align*}
(\rho)_{n \separator m} &\equiv \bra{n}\rho\ket{m}	\\
(\Lambda_{\2E})_{n,i \separator m,j} &\equiv \bra{n,i}\Lambda_{\2E}\ket{m,j}.
\end{align*}

\subsection{Vectorization}
In the following derivations we will also use the notion of \emph{vectorization} of an operator. Let $A\in L(\2X)$ be a $d\times d$ matrix. We define the vectorization of $A$ to be the $d\times d$ column vector $\dket{A}\in \2X\otimes\2X$ given by stacking the columns of $A$.
Explicitly if $A = \sum_{ij} A_{ij} \ket{i}\bra{j}$ then
\beno
\dket{A} = \sum_{ij}  A_{ij} \ket{j,i}.
\eeno

\section{Constructing $\mathcal M$}
Consider a system with Hilbert space $\2X_1$, and an environment with Hilbert space $\2Y_1$. Let the system and environment initially be in a $\rho_{\textsc{se}}\in L(\2X_1\otimes\2Y_1)$. Consider the case where we first apply to this state a preparation procedure $\mathcal P = \mathcal{P}_{\textsc{s}}\otimes \mathcal I_{\textsc{e}} \in T(\2X_1\otimes\2Y_1, \2X_2,\otimes\2Y_2)$, where $\2Y_2=\2Y_1$ and $\2P_{\textsc{s}}\in T(\2X_1,\2X_2)$ acts only on the system to prepare it in a desired input state. This is followed by coupled evolution of the joint system-environment state, described by a CPTP map $\mathcal U \in T(\2X_2\otimes\2Y_2,\2X_3\otimes\2Y_3)$, see Fig.~\ref{fig:Motivation}b). 
The output is then given by
\begin{align}
\rho_s^\prime 
&= \Tr_{\2Y_3}\big[\2U(\2P_{\textsc{s}}(\rho_{\textsc{se}}))\big]  \nonumber \\
&= \Tr_{\2X_2,\2Y_2,\2Y_3}\Big[\big(\2P(\rho_{\textsc{se}})^T\otimes\id_{se}\big)\Lambda_{\2U}\Big] \nonumber \\
&=  \Tr_{\2X_2,\2Y_2,\2Y_3}\Big[\big(\Tr_{\2X_1,\2Y_1}\Big[\big(\rho_{\textsc{se}}^T\otimes\id_{\textsc{se}})\Lambda_{\2P}\big]^T\otimes\id_{\textsc{se}}\big) \Lambda_{\2U}\Big] \label{eq:m-map-evo}
\end{align}
Note that $\Lambda_{\mathcal U}$ and $\Lambda_{\mathcal P}$ have each 4 subsystem indices, which correspond to $S$-input, $E$-input, $S$-output and $E$-output, respectively.

\begin{figure}[h]
  \begin{center}
\includegraphics[width=0.50\columnwidth]{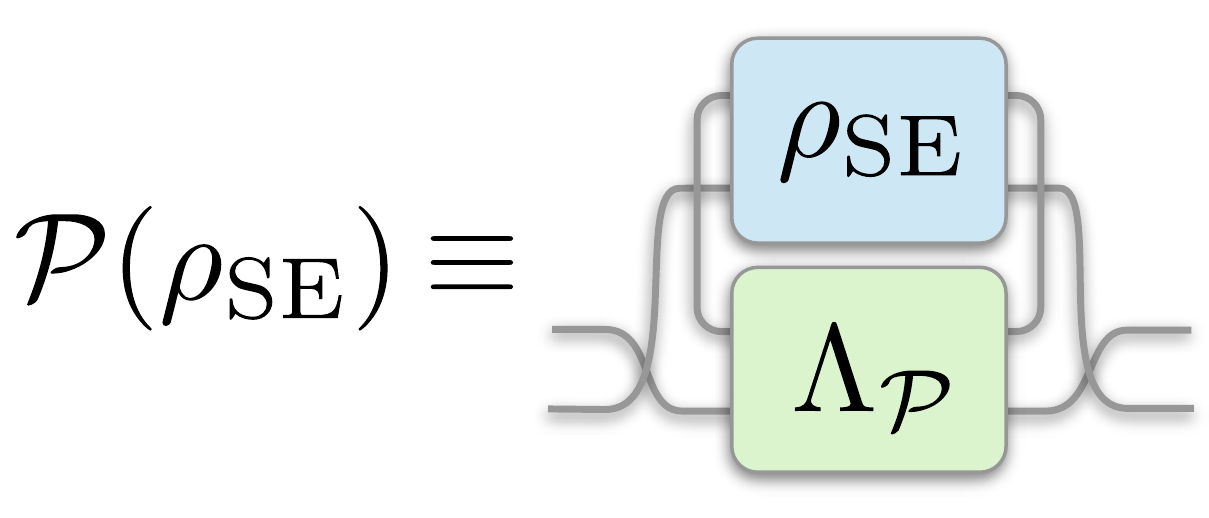}
  \end{center}
  \vspace{-8mm}
\caption{Graphical depiction of the Choi matrix representation for the state preparation procedure $\2P$ as given in Eq.~\eqref{eq:prep-map}.}
  \label{fig:tn-choi-prep}
\end{figure}

\begin{figure}[h!]
  \begin{center}
\includegraphics[width=0.50\columnwidth]{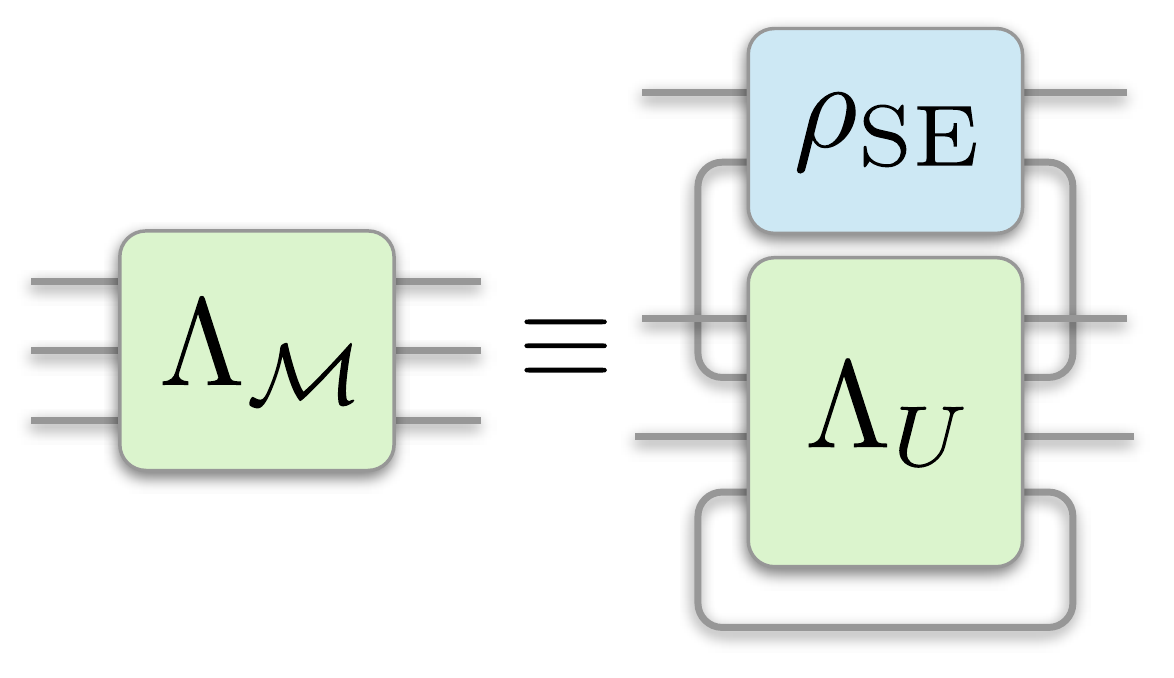}
  \end{center}
  \vspace{-8mm}
\caption{Graphical depiction of the Choi matrix for a superchannel $\2M$ for describing the reduced dynamics of a system initially correlated with its environment as given in Eq.~\eqref{eq:m-map}. Since this is a contraction of two positive semidefinite matrices -- the input density matrix, and the SE interaction $\2U$ --- it is  also a positive-semidefinite matrix. Hence the superchannel $\2M$ is always a CP-map.}
  \label{fig:tn-choi-M}
\end{figure}

In terms of index contractions
\begin{align*}
(\rho_{\textsc{se}})_{i_1i_2 \separator j_1j_2} 
	&= \bra{i_1i_2}\rho_{\textsc{se}}\ket{j_1j_2}	\\
(\Lambda_{\2U})_{i_1i_2i_3i_4 \separator j_1j_2j_3j_4} 
	&= \bra{i_1i_2i_3i_4}\Lambda_{\2U}\ket{j_1j_2j_3j_4}	\\
(\Lambda_{\2P_s})_{i_1i_2 \separator j_1j_2} 
	&= \bra{i_1i_2}\Lambda_{\2P_s}\ket{j_1j_2}\\
(\Lambda_{\2I_e})_{i_1i_2 \separator j_1j_2}
	 &= \bra{i_1i_2}\Lambda_{\2I_e}\ket{j_1j_2}\\
	&= \delta_{i_1i_2}\delta_{j_1j_2} ,
\end{align*}
we can write the action of the preparation map as
\begin{equation}
\2P(\rho_{\textsc{se}})_{i_1i_2 \separator j_1j_2}  =  \sum_{n,m}(\rho_{\textsc{se}})_{ni_2 \separator mj_2} (\Lambda_{\2P_{\textsc{s}}})_{ni_1 \separator mj_1} ,
\label{eq:prep-map}
\end{equation}
see Fig.~\ref{fig:tn-choi-prep} for the graphical representation.
The final output is thus given by
\begin{align}
(\rho^\prime)_{i_1\separator j_1} 
	&= \sum_{\mathclap{n_1,n_2, m_1 \atop m_2,z}} \2P(\rho_{\textsc{se}})_{n_1n_2 \separator m_1m_2} (\Lambda_{\2U})_{n_1n_2i_1z \separator m_1m_2j_1z}	 \label{eq:OutputIndex}\\
	&= \sum_{\mathclap{n_1,n_2,m_1,m_2,\atop m_3,x,y,z}} 
		(\rho_{\textsc{se}})_{xn_2 \separator ym_2} (\Lambda_{\2P_{\textsc{s}}})_{xn_1 \separator ym_1} (\Lambda_{\2U})_{n_1n_2i_1z \separator m_1m_2j_1z} ,\nonumber
\end{align}
which is illustrated graphically in Fig.~\ref{fig:tn-choi-derivation}.
We now define $\Lambda_{\mathcal M}$ in terms of the initial system-environment state $\rho_{\textsc{se}}$ and the interaction Choi matrix $\Lambda_{\mathcal{U}}$ as
\begin{equation}
(\Lambda_{\2M})_{i_1i_2i_3\separator j_1j_2j_3} = \sum_{n,m,l}(\rho_{\textsc{se}})_{i_1n \separator j_1m} (\Lambda_{\2U})_{i_1ni_3l\separator  j_1mj_3l} ,
\label{eq:m-map}
\end{equation}
see Fig.~\ref{fig:tn-choi-M}.

By construction, $\Lambda_{\2M}\ge 0$ if $\2U$ is CP. Note, however, that the same does not hold true for TP. If $\mathcal U$ is TP, then
\beno
\Tr_{\2X_3}[\Lambda_{\2M}] = \Tr_{\2Y_1}[\rho_{\textsc{se}}]\otimes\id_{\mathcal X_2}.
\eeno
Hence $\2M$ is TP if and only if $\Tr_{\2Y_1}[\rho_{\textsc{se}}]=\id_{\mathcal X_1}$, which is the case only for a maximally entangled initial state or if the system is initially in a completely mixed state. In all other cases different preparation procedures would lead to different overall count rates.
For a TP map $\mathcal U$, the Choi matrix for $\2M$ has normalization 
\begin{equation*}
\Tr[\Lambda_{\2M}] = \Tr[\2U(\id_{\mathcal X_1}\otimes\Tr_{\2X_1}[\rho_{\textsc{se}}])] = d_{\mathcal X_1}
\end{equation*}
The quantum superchannel $\mathcal M\in T(\2X_1\otimes\2X_2,\2X_3)$ thus takes the system-preparation procedure $\Lambda_{\mathcal P_{\textsc{s}}}\in L(\2X_1\otimes\2X_2)$ as an input and produces an output quantum state $\rho^\prime \in L(\2X_3)$ given by
\begin{align*}
\rho^\prime &= \mathcal M(\Lambda_{\mathcal P_{\textsc{s}}}) \\
&= \Tr_{\2X_1,\2X_2}[(\Lambda_{\mathcal P_{\textsc{s}}} \otimes \id_{\mathcal X_3}) \Lambda_{\mathcal U}] .
\end{align*}

\section{Tomographic reconstruction of $\2M$}
\subsection{Projective preparations}
A projective preparation procedure $\2P_{ij}\in T(\2X_1,\2X_2)$, such as the one used in our experiment, consists of an initial projection (or postselection) onto the state $\rho_i$ followed by a rotation to the state $\rho_j$. The corresponding preparation map on the system is described by the Choi matrix
\begin{equation*}
\Lambda_{\2P_{ij}} = \rho_i^\ast\otimes\rho_j ,
\end{equation*}
where $^\ast$ denotes complex conjugation.

Hence the probability for observing a count when preparing a state using $\Lambda_{\2P_{ij}}$ and then measuring the system by projecting onto a state $\rho_k$ is given by
\begin{align}
p_{ijk}
	&= \Tr\big[\rho_k^\dagger \Tr_{12}[(\rho_i^\dagger\otimes\rho_j^T\otimes\id_{\mathcal X_3})\Lambda_{\2M}]\big]	\nonumber\\
	&= \Tr\big[(\rho_i^\dagger\otimes\rho_j^T\otimes\rho_k^\dagger)\Lambda_{\2M}\big]	\nonumber\\
	&= \dbradket{\rho_i\otimes{\rho_j^\ast}\otimes\rho_k}{\Lambda_{\2M}} \nonumber\\
	&= \dbradket{\Pi_{ijk}}{\Lambda_{\2M}} ,\nonumber
\label{eq:Probs}
\end{align}
where $\Pi_{ijk} \equiv \rho_i\otimes\rho_j^\ast\otimes\rho_k$.

\subsection{Linear inversion}
Let $\{\rho_i\}_{i=1}^K$, with $K\ge d^2$ be a set of input states which span the state-space, and let $p_{ijk}$ be the probability of observing a count for preparation procedure $\2P_{ij}$, and final measurement $\rho_k$. We may reconstruct $\Lambda_{\2M}$ from observed estimates of $p_{ijk}$ via linear inversion as follows.

Define a vector $\ket{p}$, and matrices $S$ and $W$ by
\begin{align*}
\ket{p} &= \sum_{i,j,k=1}^K p_{ijk}\ket{i,j,k}	\\
S 	&= \sum_{i,j,k=1}^K \ket{i,j,k}\dbra{\Pi_{ijk}}	\\
W	&= \sum_{i,j,k=1}^K w_{ijk}\ketbra{i,j,k}.
\end{align*}
where $w_{ijk}\ge 0$ are weights which specify the relative importance of various projectors. The linear inversion estimate of $\Lambda_{\2E}$ is given by the weighed least-squares fit
\begin{equation*}
\hat \Lambda_{\2E} = \mbox{argmin}_{\Lambda_{\2E}} \| WS\dket{\Lambda_{\2E}}-W\ket{p} \|_2
\label{eq:least-sq}
\end{equation*}
where $\|\cdot\|_2$ is the Euclidean vector norm. This has the analytic solution
\begin{align}
\dket{\hat \Lambda_{\2E}} 
	&= (S^\dagger W^\dagger W S)^{-1} S^\dagger W^\dagger W\ket{p} 
	\label{eq:lin-inv}\\
	&= \sum_{i,j,k}w_{ijk}^2 p_{ijk}\left(\sum_{l,m,n}w_{lmn}^2 \dketdbra{\Pi_{lmn}}{\Pi_{lmn}}\right)^{-1} \dket{\Pi_{ijk}}\nonumber
\end{align}

In the case where we set uniform weights $w_{ijk}{=}1$, this is often expressed in terms of a dual set $\{D_{ijk}\}$ defined by
\begin{equation*}
\dket{D_{ijk}}	
	= \left(\sum_{l,m,n}\dketdbra{\Pi_{lmn}}{\Pi_{lmn}}\right)^{-1}\dket{\Pi_{ijk}},
\end{equation*}
and then \eqref{eq:lin-inv} reduces to
\begin{equation*}
\hat \Lambda_{\2E} = \sum_{i,j,k} p_{ijk} D_{ijk}
\end{equation*}

Linear inversion is the standard technique used in QPT for tomographic reconstruction of the Choi matrix of an unknown quantum process, under the assumption of vanishing initial system-environment correlations. However, it is susceptible to statistical errors due to finite count rates, (non-uniform) losses in the experiment, or other random fluctuations during data acquisition. In the presence of these errors the result of QPT may be an unphysical (i.e.\ non-CP) description of the process. More sophisticated techniques, most notably maximum-likelihood estimation~\cite{Hradil1997,James2001} and Bayesian mean estimation~\cite{Blume-Kohout2010a} have been developed to address these issues. By imposing additional constraints, these methods aim to reconstruct the \emph{physically meaningful} (i.e.\ CP) description that is the closest fit to the observed experimental data for a given objective function.

For standard QPT this restriction, however, prevents the correct reconstruction of a genuine non-CP map, which might arise for the reduced dynamics of the system in the presence of initial correlations. In particular, conventional QPT is not able to distinguish such situations from the apparent non-CP dynamics due to statistical errors. 
The superchannel $\mathcal M$ in contrast, fully takes the effect of state preparation into account. It is thus always a CP-map and the additional constraint is justified even in the presence of initial correlations.

\subsection{Maximum likelihood estimation}
\label{sec:SI_MLE}
Maximum likelihood estimation (MLE) is the constrained optimization problem
\begin{align}
&\text{minimize}\quad \| WS\dket{\Lambda_{\2E}}-W\ket{p} \|_2
\nonumber\\
&\text{subject to: } \quad\Lambda_{\2E} \ge 0, \quad \Tr[\Lambda_{\2E}] = d
\label{eq:MLEopt}
\end{align}
Since the objective function $\| \cdot \|_2$ in Eq.~\eqref{eq:MLEopt} is convex, and the constrains are semidefinite, the problem can be solved numerical as a \emph{semidefinite program} (SDP). We implemented this SDP using the CVX optimization package in MATLAB~\cite{cvx}. 
For the choice of weights $w_{ijk}$ we assumed a normal approximation for the distribution of the observed probabilities $p_{ijk}$ so that 
\beno
w_{ijk} = \sqrt{\frac{N_j}{p_j(1-p_j)}}
\eeno

The objective function is thus equivalent to 
\beno
\mbox{minimize}\quad  \sum_{i,j,k} 
	\frac{N_j(\Tr[\Pi_{ijk}^\dagger \Lambda_{\2E}]-p_{ijk})^2}{p_j(1-p_j)}
\eeno
Note that this approximation is not well-defined for $p_j = 0,1$. To overcome this and other common issues associated with zero probabilities in MLE~\cite{Blume-Kohout2010} we use \emph{hedged} MLE, defining our observed probabilities as
\beno
p_{ijk} = \frac{n_{ijk} + \beta}{N_{jik}+K\beta}
\eeno
where $n_{ijk}$ is the observed number of counts out of $N_{ijk}$ trials for a given projector $\Pi_{ijk}$, $K$ is the number of possible outcomes of a given measurement, and $\beta$ is a small hedging parameter. Since our experiment had good counting statistics ($N_{ijk} > 5000$) we used a small value of $\beta = 0.1$. We note, however, that the reconstruction is not particularly sensitive to the value of $\beta$.

Since $N_{ijk}$ is unknown a priori we define it for our experiment by totalling the observed counts for measurement configurations that sum to identity. Since the second index of $\Pi_{ijk}$ corresponds to the rotated state for the initial projective preparation procedure, only the first and third indices correspond to true measurements and so we have $K=4$. E.g.\ if $\rho_0 = \ketbra{H}, \rho_1 = \ketbra{V}$, then  $\sum_{i,k=0,1}\rho_i\otimes\rho_k= \id\otimes\id$.

\section{Additional experimental data}
\begin{figure}[h!]
  \begin{center}
\includegraphics[width=\columnwidth]{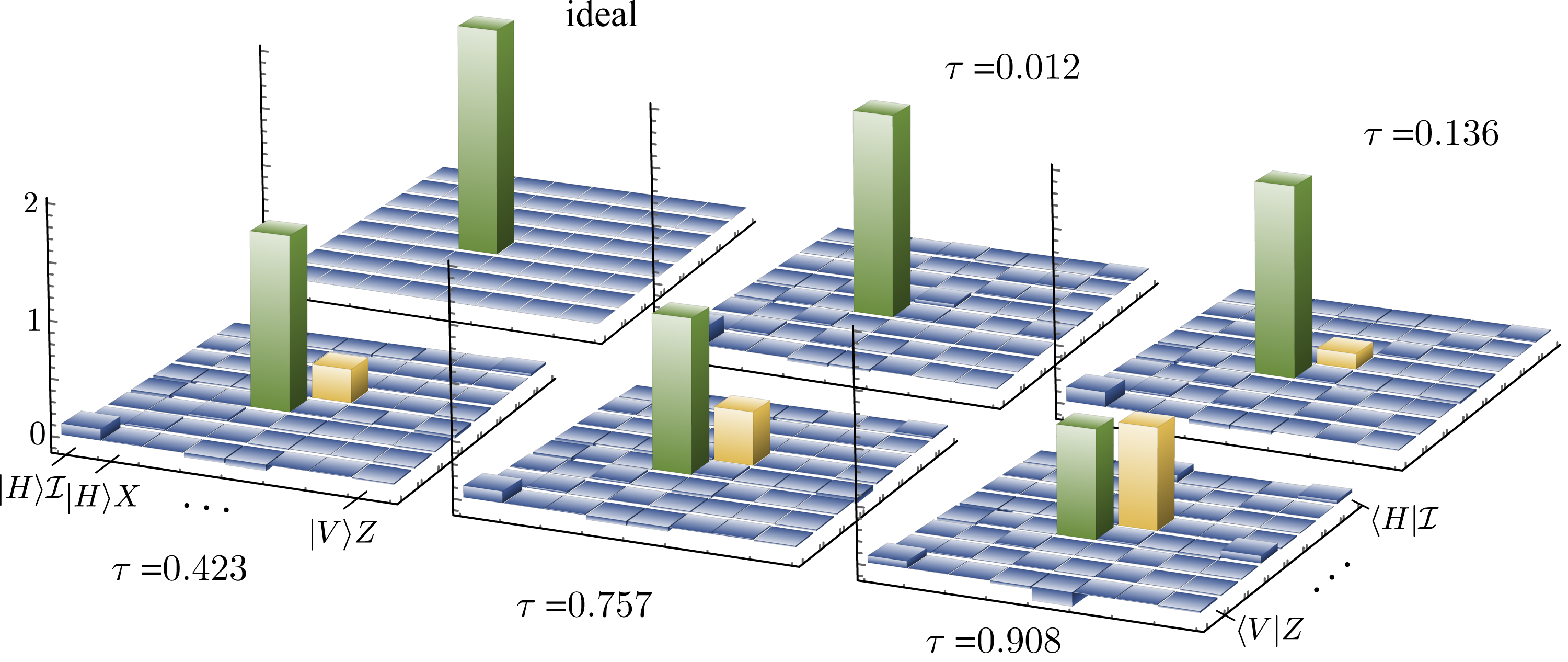}
  \end{center}
  \vspace{-8mm}
\caption{Real parts of $\Lambda_{\mathcal{M}}$ for an indented $U_{\textsc{s}}=Z$ operation on the system in the ideal, uncorrelated case and experimentally for increasing strength of initial correlations. The matrices $\Lambda_{\mathcal M}$ are shown in a compound polarization-Pauli basis, as in Fig.~\ref{fig:results_map}. The imaginary parts are smaller by at least two orders of magnitude and thus not shown.}
  \label{fig:results_CZ}
\end{figure}

\begin{figure}[h!]
  \begin{center}
\includegraphics[width=\columnwidth]{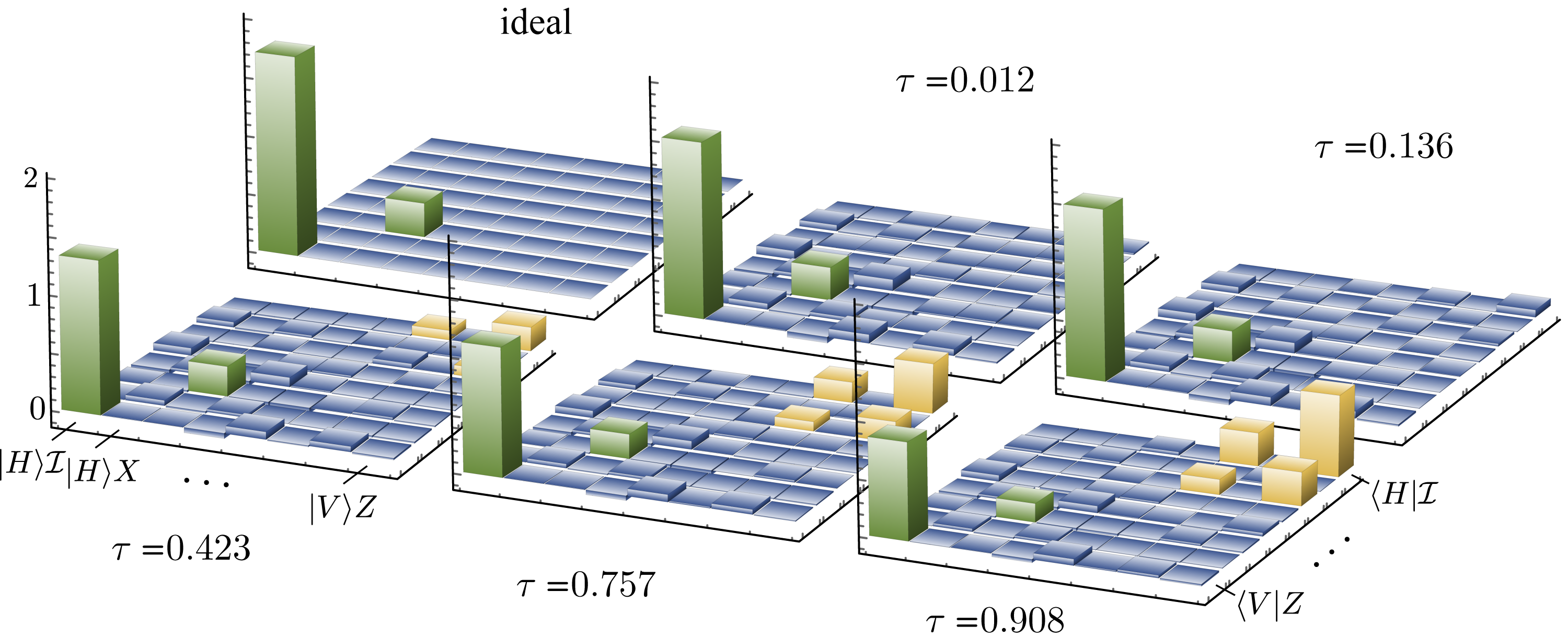}
  \end{center}
  \vspace{-8mm}
\caption{Real parts of $\Lambda_{\mathcal{M}}$ for an indented $U=R_{\textsc{y}}Z$ operation on the system in the ideal, uncorrelated case and experimentally for increasing strength of initial correlations. The matrices $\Lambda_{\mathcal M}$ are shown in a compound polarization-Pauli basis, as in Fig.~\ref{fig:results_map}.}
  \label{fig:results_CR_Re}
\end{figure}

\begin{figure}[h!]
  \begin{center}
\includegraphics[width=\columnwidth]{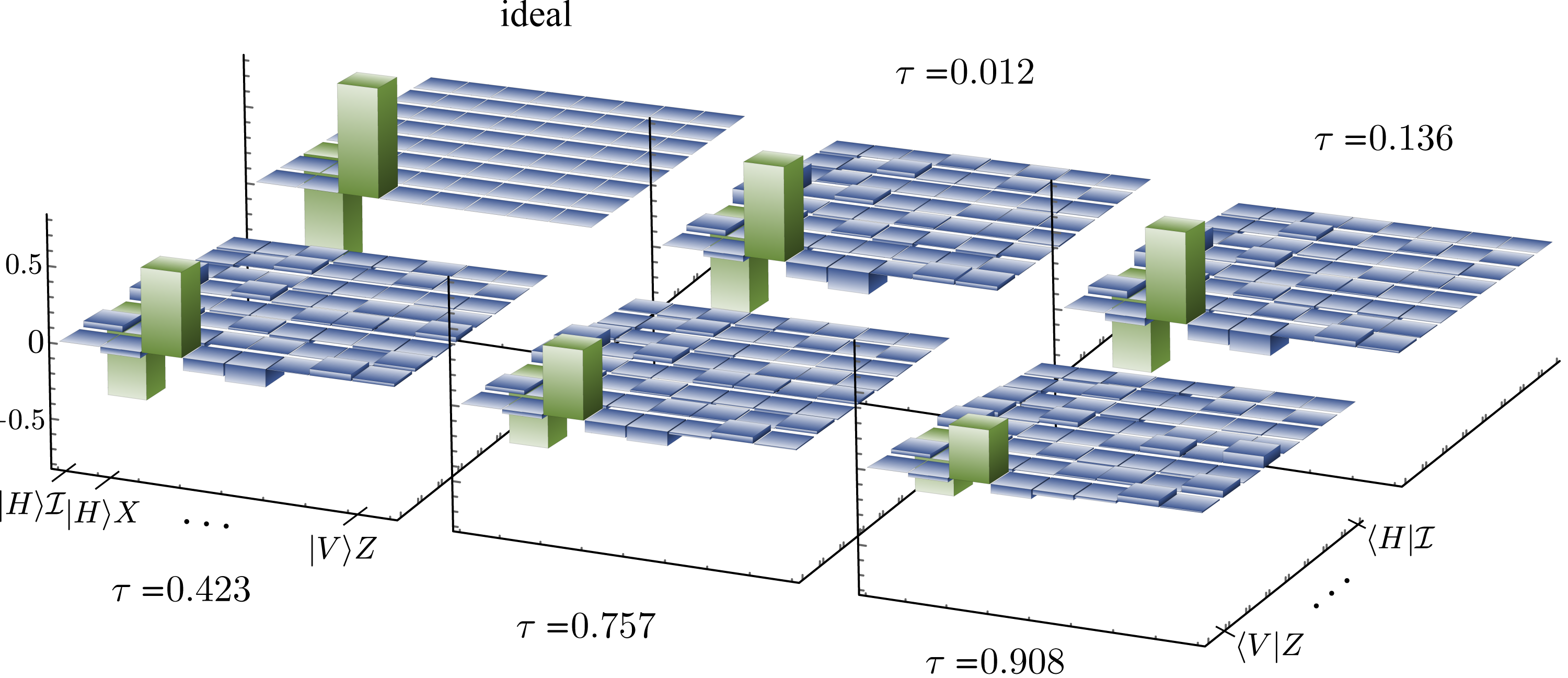}
  \end{center}
  \vspace{-8mm}
\caption{Imaginary parts of $\Lambda_{\mathcal{M}}$ for an indented $U_{\textsc{s}}=R_{\textsc{y}}Z$ operation on the system in the ideal, uncorrelated case and experimentally for increasing strength of initial correlations. The matrices $\Lambda_{\mathcal M}$ are shown in a compound polarization-Pauli basis, as in Fig.~\ref{fig:results_map}.}
  \label{fig:results_CR_Im}
\end{figure}

Reconstructed superchannels for nominal system unitaries $U_{\textsc{s}}=Z$ are shown in Fig.~\ref{fig:results_CZ} and for $U_{\textsc{s}}=R_{\textsc{y}}Z$ in Fig.~\ref{fig:results_CR_Re} and~\ref{fig:results_CR_Im}.

\section{Graphical calculus}
\begin{figure}[h!]
  \begin{center}
\includegraphics[width=\columnwidth]{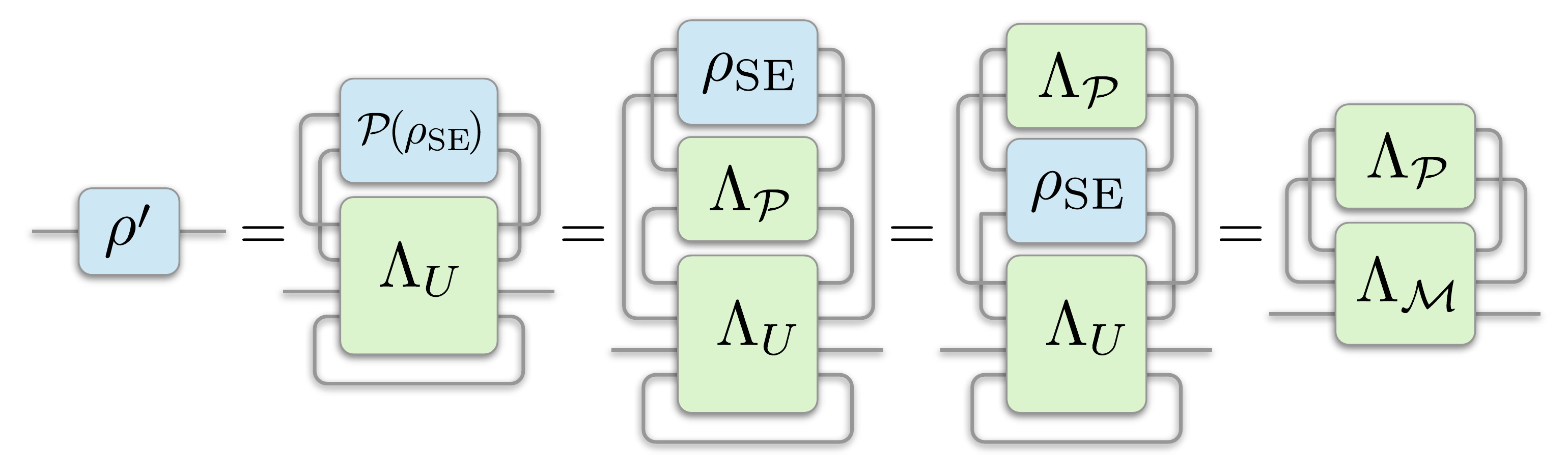}
  \end{center}
  \vspace{-8mm}
\caption{Graphical derivation of the superchannel $\2M$ for describing the reduced dynamics of a system initially correlated with the environment. Since all the tensor wires are contracted we may essentially interchange the position of $\rho_{\textsc{se}}$ and $\Lambda_{\2P}$ to make the preparation procedure the effective input state. The resulting tensor network contracting of $\rho_\textsc{se}$ and $\Lambda_{\2U}$ then defines $\Lambda_{\2M}$ equivalently to the index contraction given in Eq.~\eqref{eq:m-map-evo}.}
  \label{fig:tn-choi-derivation}
\end{figure}

When contracting complex sequences of tensors it can often be useful to employ diagrammatic representations of these contractions in terms of \emph{tensor networks}. In particular a graphical calculus for CP-maps and open quantum systems was presented in Ref.~\cite{Wood2011}. 
These graphical techniques can greatly facilitate complete index manipulations, such as Eq.~\eqref{eq:OutputIndex} and Eq.~\eqref{eq:m-map}, that can quickly become cumbersome when using index contraction notation. In the present case, the use of this notation is especially useful for the construction of the superchannel $\2M$, as illustrated in Fig.~\ref{fig:tn-choi-derivation}. In this section we briefly summarize the key features of these graphical techniques and illustrate their use for simplifying manipulations of $\2 M$

\begin{figure}[h!]
  \begin{center}
\includegraphics[width=0.8\columnwidth]{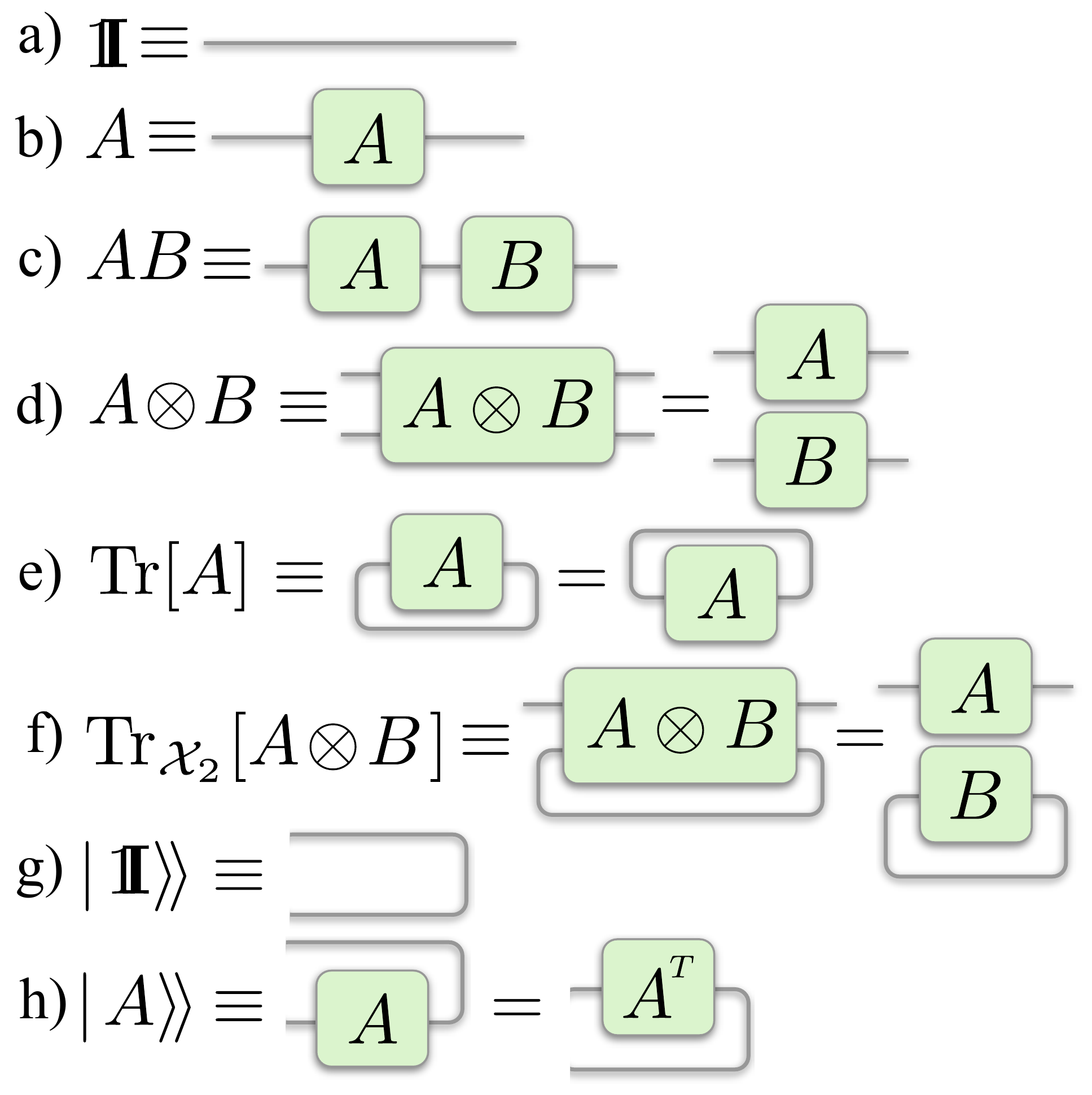}
  \end{center}
  \vspace{-8mm}
\caption{\textbf{a)} A straight wire corresponds to an identity matrix.
\\\textbf{b)} A box with two wires corresponds to a matrix (rank-2) tensor. 
\\\textbf{c)} Matrix multiplication (tensor contraction) is represented by connecting the ``wires'' corresponding to the indices to be summed over.
\\\textbf{d)} Tensor products are represented by vertical composition. The rank of the resulting tensor is the sum of the tensor ranks.
\\\textbf{e)} The trace operation for square matrices is represented by contracting the two matrix indices with each other.
\\\textbf{f)} The partial trace operation for composite square matrices is represented by contracting the two matrix indices of the corresponding subsystems to be traced over.
\\\textbf{g)} The vectorized identity operator is equivalent to the unnormalized bell state $\sum_i \ket{i}\otimes\ket{i}$.
\\\textbf{h)} Column vectorization of a matrix is represented by ``bending'' the right wire upwards to form a vector.
``Sliding'' a matrix around a bell-state results in the transposition of the matrix.}
  \label{fig:tn-basics}
\end{figure}

The basic components of the graphical calculus are illustrated in Fig.~\ref{fig:tn-basics}. A rank-$r$ tensor will have $r$ free \emph{wires}, where the orientation of these wires depicts whether the tensor is a vector (all wires going to the left), dual vector (all wires going to the right), matrix (wires going to the left and right) or a scalar (no free wires). These diagrams are oriented to read from right to left to mimic the underlying algebraic operations which is the opposite convention of standard quantum circuits. Connecting wires corresponds to contracting the underlying indices (such as matrix multiplication or trace operations). Vertical composition represents tensor products, and operations such as vectorization have natural representations by bending around wires of matrices to turn them into vectors (for more details on these techniques see~\cite{Wood2011}).

\begin{figure}[h!]
  \begin{center}
\includegraphics[width=0.85\columnwidth]{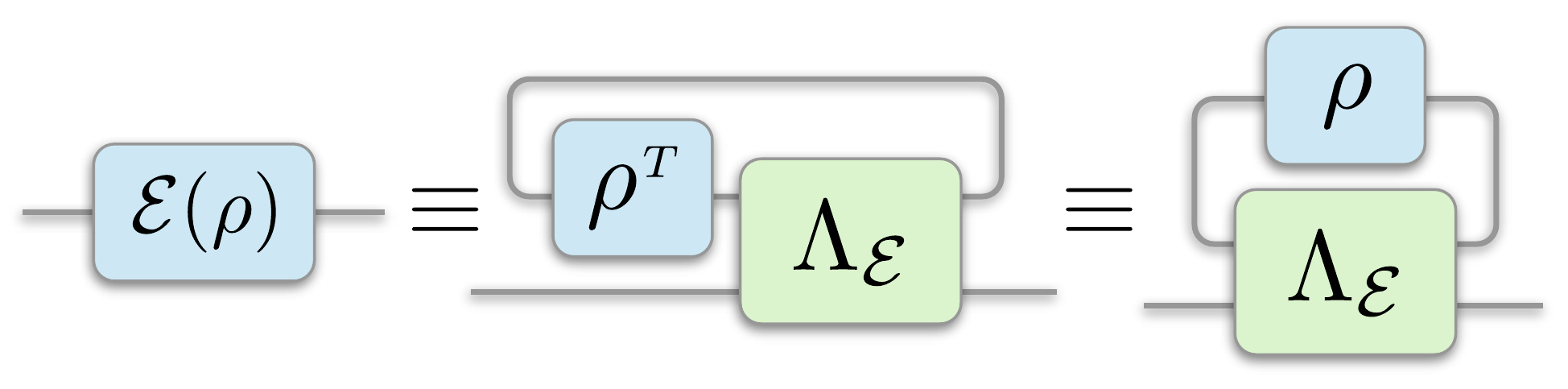}
  \end{center}
  \vspace{-8mm}
\caption{Graphical depiction of the Choi matrix representation for the evolution of a density matrix $\rho$ by a channel $\2E$ as given in Eq.~\eqref{eq:choievo}, where the closed wire on the first subsystem denotes the partial trace $\Tr_{\2X_1}[(\rho^T\otimes\id)\Lambda_{\2E}]$.}
  \label{fig:tn-choi}
\end{figure}

\begin{figure}[h!]
  \begin{center}
\includegraphics[width=0.55\columnwidth]{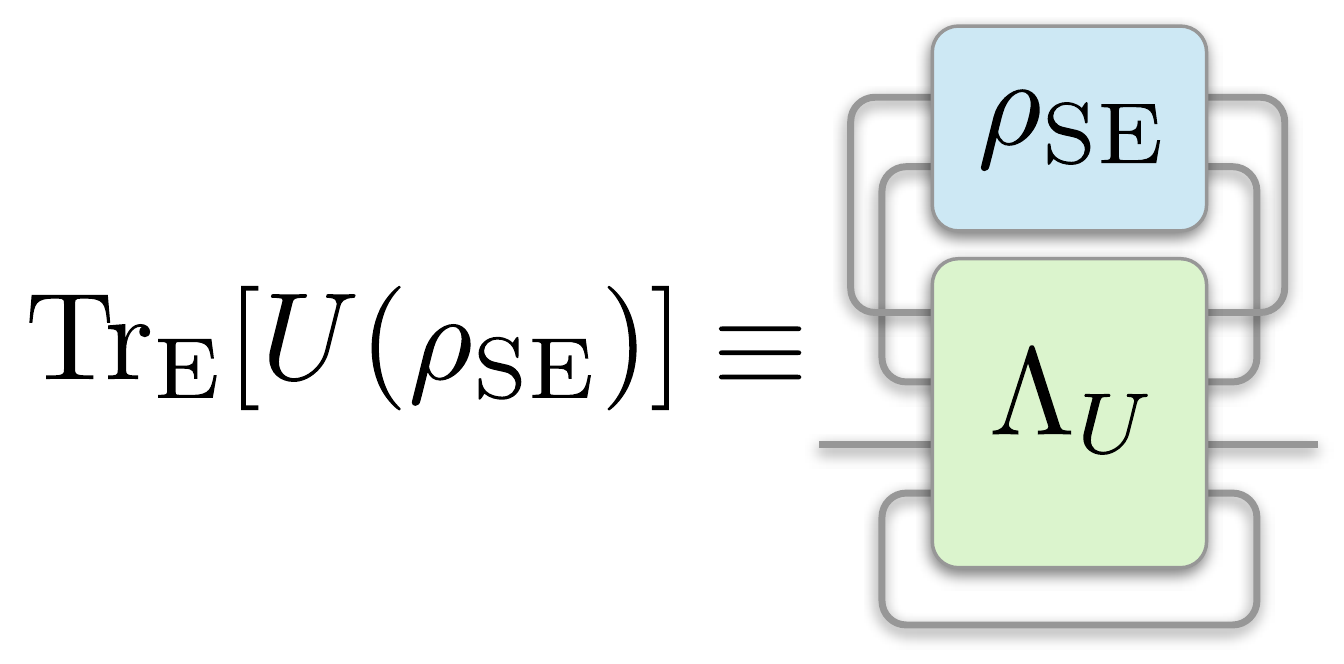}
  \end{center}
  \vspace{-8mm}
\caption{Graphical depiction of the Choi matrix representation for the evolution of a joint system-environment density matrix $\rho_\textsc{se}$ by the joint unitary channel $U$. The final partial trace over the environment degrees of freedom correspond to the closer wire over the forth indices of $\Lambda_U$.}
  \label{fig:tn-choi-se}
\end{figure}

Equation~\eqref{eq:choievo} for describing the evolution of a density matrix $\rho\in\2L(\2X)$ in terms of the Choi matrix $\Lambda_{\2E}$ for a channel $\2E\in T(\2X)$ is illustrated in Fig. \ref{fig:tn-choi}. In the case where we explicitly include the environment, the joint state $\rho_\textsc{se}$ evolves as in Fig.~\ref{fig:tn-choi-se}. Here we use different colors to distinguish Choi matrices (green) from state matrices (blue), rather than using them to denote a summation convention as in Ref.~\cite{Wood2011}.

Even for small systems of only two qubits index manipulations such as in Eq.~\eqref{eq:OutputIndex} are cumbersome and can be simplified by using the graphical approach as illustrated in Fig.~\ref{fig:tn-choi-derivation} for the construction of the superchannel $\mathcal{M}$.

\end{document}